\documentclass[aps,preprintnumbers,amsmath,twocolumn,groupedaddress,superscriptaddress,notitlepage,nofootinbib,prd]{revtex4-1}
\usepackage[utf8]{inputenc}
\usepackage{indentfirst}
\usepackage{amssymb}
\usepackage{amsmath,latexsym}
\usepackage{graphicx}
\usepackage{float}
\usepackage[margin=1.0in]{geometry}
\usepackage[mathscr]{euscript}
\usepackage{color}
\usepackage[dvipsnames]{xcolor}
\usepackage{soul}
\usepackage{slashed}
\usepackage{feynmf}
\usepackage{braket}
\usepackage{ulem}
\usepackage{comment}

\DeclareMathOperator{\tr}{tr}

\usepackage{dsfont}
\newcommand{\bbid}{\mathds{1}}
\newcommand{\beq}{\begin{equation}}
\newcommand{\eeq}{\end{equation}}
\newcommand{\bea}{\begin{eqnarray}}
\newcommand{\eea}{\end{eqnarray}}

\long\def\beqs#1\eeqs{\beq\begin{split} #1 \end{split}\eeq}

\newcommand{\bo}{\mathbf}

\definecolor{MyRed}{RGB}{153,0,13}

\usepackage[colorlinks=true,backref=true,linktocpage=true,citecolor=MyRed,urlcolor=MyRed,linkcolor=MyRed,pdfpagemode=UseOutlines]{hyperref}

\hypersetup{%
  bookmarksnumbered=true,
  pdftitle = {},
  pdfsubject = {},
  pdfauthor = {},
  pdfkeywords = {}
}

\usepackage{microtype}

\usepackage{pgfplotstable,array,siunitx}
\usepackage{multirow}
\usepackage{booktabs}

\usepackage{tikz}

\preprint{INT-PUB-20-030}

\newcommand{\fig}[1]{Fig.~\ref{#1}}

\begin{document}
\title{Qubitization strategies for bosonic field theories}

\author{Andrei Alexandru}
\email{aalexan@gwu.edu}
\affiliation{Department of Physics,
The George Washington University, Washington, DC  20052}
\affiliation{Department of Physics,
University of Maryland, College Park, MD 20742}

\author{Paulo F. Bedaque}
\email{bedaque@umd.edu}
\affiliation{Department of Physics,
University of Maryland, College Park, MD 20742}

\author{Andrea Carosso}
\email{acarosso@gwu.edu}
\affiliation{Department of Physics,
The George Washington University, Washington, DC  20052}

\author{Michael J. Cervia}
\email{cervia@gwu.edu}
\affiliation{Department of Physics,
The George Washington University, Washington, DC  20052}
\affiliation{Department of Physics,
University of Maryland, College Park, MD 20742}

\author{Andy Sheng}
\email{asheng@umd.edu}
\affiliation{Department of Physics,
University of Maryland, College Park, MD 20742}

\preprint{}

\date{\today}
\pacs{}

\begin{abstract}

Quantum simulations of bosonic field theories require a truncation in field space to map the theory onto finite quantum registers. Ideally, the truncated theory preserves the symmetries of the original model and has a critical point in the same universality class. In this paper, we explore two different truncations that preserve the symmetries of the 1+1-dimensional $O(3)$ non-linear $\sigma$-model --- one that truncates the Hilbert space for the unit sphere by setting an angular momentum cutoff and a {\it fuzzy sphere} truncation inspired by non-commutative geometry. We compare the spectrum of the truncated theories in a finite box with the full theory. We use {\it open} boundary conditions, a novel method that improves on the correlation lengths accessible in our calculations. We provide evidence that the angular-momentum truncation fails to reproduce the $\sigma$-model and that the anti-ferromagnetic fuzzy model agrees with the full theory.

\end{abstract}

\maketitle

\section{Introduction}

Recent developments in quantum computing promise to open new areas of investigations for quantum field theories (QFTs). Of particular interest are real-time evolution and properties of high density matter, where simulations on classical computers are very challenging due to the sign problem.

Typically to study a QFT numerically, the theory is regularized using a lattice discretization with the fields sampled on a regular spatial grid. The physical degrees of freedom at every site are then mapped onto qubits. If we are interested in real-time dynamics for this system, the unitary evolution $\exp(-iHt)$ can be implemented using a sequence of quantum gates. For a local Hamiltonian, this method leads to a circuit that has a fixed depth per unit time, and the number of gates increases only linearly with the number of points in the grid. Compare this scaling with the classical computer methods for which the numerical cost increases exponentially with the size of the grid, since the size of the matrix representing $H$ grows exponentially with the volume. Quantum computers have a definite advantage here. 

For this program to work, one must first deal with a subtle complication. While for fermionic fields the Hilbert space at each site is finite and can be mapped faithfully onto a set of qubits, for bosonic fields this Hilbert space is infinite and mapping it to qubits requires a truncation ({\it qubitization}\/). To recover the physical results we need to perform a double limit, one to remove the lattice regulator (the {\it continuum limit}) and another to remove the bosonic space truncation. This double limit is not only cumbersome and expensive in the number of qubits and gates required, it is, in some contexts, not even possible since the truncation cannot be made arbitrarily fine~\cite{Alexandru:2019nsa,Alexandru:2021jpm}.

In a previous study we proposed to rely on universality to circumvent the double limit problem~\cite{Alexandru:2019ozf}. To understand the proposal, recall that to perform the continuum limit for lattice QFTs, we tune the system to criticality where the correlation lengths in lattice units go to infinity. The discretization artifacts vanish, and the continuum limit results are independent of the particular discretization employed. We proposed using truncated lattice QFTs that have critical points in the same universality class as the original theory. For such models, the continuum limit automatically reproduces the full results. 

The design principle for such truncations is 
the belief that two theories differing only at short distances and possessing the same symmetries lead to the same continuum limit.
This principle has only suggestive power; preserving the symmetries is neither necessary nor sufficient to have the correct continuum limit.
What need to be verified on a case-by-case basis are that:
the truncated theory has a critical point, and simulations close to the relevant critical point recovers the class of continuum results~\cite{Bhattacharya:2020gpm,Alexandru:2021xkf}.

In this paper, we explore this idea using the 1+1-dimensional $O(3)$ $\sigma$-model. This model is asymptotically free with a mass gap. The usual representation in Euclidean time involves fields that take values on the unit 2-sphere and the action is invariant under global $O(3)$ rotations in this space. We mention in passing that one possible truncation for this model involves sampling the field values at a finite subset of points~\cite{Patrascioiu:1997ds, Hasenfratz:2000sa}, breaking the $O(3)$ symmetry. Whether this truncation has a critical point that reproduces the full theory results is still debated~\cite{Caracciolo:2001jd,Hasenfratz:2001iz}. 

In Sec.~\ref{sec:models}, we outline two truncation strategies preserving $O(3)$ invariance that we will analyze in this paper. A widely used proposal is to truncate the Hilbert space of complex wave functions on the sphere using an angular momentum cutoff $\ell_{\text{max}}$~\cite{Bruckmann:2019, liu2021qubit}.\footnote{In fact, Ref.~\cite{Bruckmann:2019} provides evidence that the standard lattice discretization of the $\sigma$-model can be reproduced in the limit of $\ell_{\text{max}} \rightarrow \infty$.} 
Here, we study the truncated model with $\ell_{\text{max}} = 1$, which requires the same number of qubits per site as the other proposal we analyze, the {\it fuzzy sphere} truncation~\cite{Alexandru:2019nsa}, inspired by ideas from non-commutative geometry~\cite{Madore:1991bw}. 

In Sec.~\ref{sec:methods}, we describe how to compute the low-lying spectrum for the truncated Hamiltonians, employing the matrix-product state (MPS) ansatz with open boundary conditions, which can better accommodate MPS calculations. We then describe how to extract the corresponding energies in Monte Carlo simulations of the full $\sigma$-model with open boundaries.


In Sec.~\ref{sec:results}, for both truncations, we compare the energy of the lowest states in a finite volume with the full $\sigma$-model results. We show that the commonly used angular momentum truncation does not reproduce the $\sigma$-model spectrum (at least for $\ell_\text{max}=1$), in contrast with the {\it anti-ferromagnetic} fuzzy model, which agrees with the full theory results up to the largest correlation length reached, $\xi\approx 66$~lattice units. This check improves on our previous work~\cite{Alexandru:2021xkf}, by showing that the qubitization agrees more deeply in the ultraviolet (i.e., for energy scales much larger than the mass gap) with the full theory. Finally, in Sec.~\ref{sec:conclude}, we discuss how our analysis of these models may assist in the general effort to successfully qubitize bosonic gauge theories.

\section{Truncated Models}
\label{sec:models}

The action of the continuum $O(3)$ $\sigma$-model is
\begin{equation}
    \label{action} S_{\sigma} = \frac{1}{2g^2} \int \mathrm{d}x \,\mathrm{d}t  \;\partial_{\mu}\bo{n}(x,t) \cdot \partial^{\mu} \bo{n}(x,t),
\end{equation}
where the field variables take values on a sphere ($\mathcal S^2$), i.e., $\bo{n} = (n_1, n_2, n_3)$ satisfies $n_1^2 + n_2^2 + n_3^2 = 1$, and $g$ is the coupling. For simulations on a quantum computer, however, we use the Hamiltonian of the discretized theory:
\begin{multline}
    H_{\sigma} = \label{genham} \sum_{x} \Big[ -\frac{g^2}{2}\nabla^2(x) - \frac{1}{g^2 a^2} \bo{n}(x+1)\cdot \bo{n}(x) \Big],
\end{multline}
where $-\nabla^2$ is the Laplace-Beltrami operator on $\mathcal{S}^2$ and $a$ is the lattice spacing. The full Hillbert space at each lattice site 
is the space of complex functions on a sphere, which is infinite-dimensional.
In order to represent this model on finitely many qubits, we need to truncate the local Hilbert space. 

\subsection{Angular Momentum Truncation}

A standard way to truncate the Hilbert space of the model is to expand functions on the sphere in terms of spherical harmonics and set an angular momentum cutoff, $\ell_{\text{max}}$ \cite{Bruckmann:2019, liu2021qubit}. States for one site are then of the form
\begin{equation}
    \Psi(\theta,\phi) = \sum_{\ell=0}^{\ell_\mathrm{max}}\sum_{m=-\ell}^{\ell} \psi_{\ell m}Y_{\ell}^m(\theta, \phi)
\end{equation}
where $(\theta,\phi)$ are the spherical angles corresponding to $\bo{n}$, $Y_{\ell}^{m}(\theta, \phi)$ are the spherical harmonics on $\mathcal{S}^2$, and $\psi_{lm}$ are the expansion coefficients. The complete set of functions on the unit sphere is recovered as we take the limit $\ell_\mathrm{max} \rightarrow \infty$. This method of truncation is easily generalizable
not only to 
any $O(N)$ $\sigma$-model,
but also to $\sigma$-models defined on other manifolds, like group manifolds and homogeneous spaces~\cite{Zohar:2014qma}. 
For the $O(3)$ $\sigma$-model we discuss here, we consider the lowest nontrivial truncation $\ell_\mathrm{max} = 1$. We choose an ordered basis $\{|\mathcal{Y}_i\rangle\} = \{Y_{0}^{0}, Y_{1}^{-1}, Y_{1}^{0}, Y_{1}^{1}\}$ for the resulting four-dimensional local Hilbert space at each site. Thus, to represent this model on a quantum computer, two qubits are required for each lattice site. 

Because the spherical harmonics are eigenfunctions of the Laplace-Beltrami operator, 
\begin{equation}
  -\nabla^2Y_{\ell}^{m} = \ell(\ell+1)Y_{\ell}^{m}
\end{equation}
we can easily write the action of $-\nabla^2/2$ on the basis $|\mathcal{Y}_i\rangle$; the kinetic term in the Hamiltonian is $H_0^c = \text{diag}(0, 1, 1, 1)$. The action of $\bo{n}(x)$, restricted to the subspace spanned by $\mathcal{Y}$, is represented by matrices $y_k$ ($k = 1, 2, 3$), the matrix elements of which are defined as
\begin{equation}
    \label{matelement} (y_{k})_{ab} = \langle \mathcal{Y}_a |n_k|\mathcal{Y}_b\rangle = \int \mathrm{d}\Omega\:\mathcal{Y}^{*}_a(\theta, \phi) n_k \mathcal{Y}_b(\theta,\phi),
\end{equation}
where the integral is over $\mathcal S^2$; $\int \mathrm{d}\Omega = \int_{0}^{2\pi}\int_{0}^{\pi}\sin(\theta)\:\mathrm{d}\theta\,\mathrm{d}\phi$. We note that $n_k$, the coordinates of the field variables, are linear combinations of $\ell = 1$ spherical harmonics, implying the $(y_{k})_{ab}$ can be obtained from Clebsch-Gordan coefficients. Higher angular momentum states which result from the addition of two $\ell=1$ states are dropped in order to remain in the Hilbert space spanned by the $\ell = 0, 1$ states. Explicitly, the $y_k$ are
\begin{multline}
    y_1 = \frac{1}{\sqrt{6}}
    \begin{pmatrix}
    0&1&0&-1\\
    1&0&0&0\\
    0&0&0&0\\
    -1&0&0&0
    \end{pmatrix}, \\
    y_2 = \frac{i}{\sqrt{6}}
    \begin{pmatrix}
    0&-1&0&-1\\
    1&0&0&0\\
    0&0&0&0\\
    1&0&0&0
    \end{pmatrix},\\
    y_3 = \frac{1}{\sqrt{3}}
    \begin{pmatrix}
    0&0&1&0\\
    0&0&0&0\\
    1&0&0&0\\
    0&0&0&0
    \end{pmatrix}.
\end{multline}
Putting the pieces together into Eq.~\eqref{genham}, the resulting truncated theory Hamiltonian for $N$ sites is:
\begin{equation}
    \label{Hamiltonian} H_{\ell_\mathrm{max}=1} = \eta g^2\sum_{x=1}^N H^{c}_{0}(x) \pm \frac{\eta}{g^2}\sum_{x=1}^{N-1}\sum_{k=1}^{3} y_{k}(x) y_{k}(x+1),
\end{equation}
where the first term is a single-site kinetic term and the second is a nearest-neighbor interaction. We stress that here we work with open boundary conditions, in which we do not include the wrap-around link $y_{k}(N) y_{k}(1)$ in the second term. For $g^2 > 0$, the relative sign~$\pm$ between the two terms corresponds to anti-ferromagnetic/ferromagnetic coupling. Also, we introduce a new useful parameter $\eta>0$, which sets an overall scale for the Hamiltonian, the tuning of which will be discussed in Sec.~\ref{sec:param_tuning}. 

The Hamiltonian presented in Eq.~\eqref{Hamiltonian} is symmetric under $O(3)$ rotations, as is the $\sigma$-model. Because the relative sign between the two terms of the Hamiltonian does not affect the $O(3)$ symmetry of the truncated model, we should consider studying the universality of both the anti-ferromagnetic and the ferromagnetic phases. However, for an even $N$ number of sites in the system, there exists a mapping between the two phases that implies they have identical spectra and similar eigenstates (at some fixed $g^2$), which we observe numerically. To define the mapping, we introduce a global operator
\begin{equation}
    \label{symmetry} \mathcal{O} = \bigotimes^{N/2}_{n=1} (U_{2n-1} \otimes \bbid_{2n}),
\end{equation}
where the operator $U$ is applied at every other site and has the property: 
\begin{equation}
    \label{condition} U H_0^c U^\dagger = H_0^c \quad\text{and}\quad U y_k U^{\dagger} = -y_k
\end{equation} for all $k = 1, 2, 3$.  In the case of the Hamiltonian in Eq.~\eqref{Hamiltonian}, it is easy to see that such an $U$ exists, we can take, for example $U = \text{diag}(1, -1, -1, -1)$. This argument can be extended to any $\ell_\text{max}$, so this equivalence between the ferromagnetic and anti-ferromagnetic couplings remains valid.


\subsection{Fuzzy Sphere Truncation}

An alternative way of truncating the field space of the $\sigma$-model is to replace $\mathcal{S}^2$ by a fuzzy sphere~\cite{Alexandru:2019ozf} --- a non-commutative approximation of the sphere \cite{Madore:1991bw,deWit:1988wri}. The coordinates of the sphere themselves are mapped to non-commuting matrices: $n_k \mapsto J_k$ and angular momentum operators to commutators: 
$-i\varepsilon_{ijk}n_i\nabla_j \mapsto [ J_k, \bullet]$. 
If we take $J_i$ as the generators of spin-$j$ irreducible representation of $SU(2)$, the spectrum of these non-commuting operators provides an approximation of the action of the original operators, which becomes exact in the limit $j \rightarrow \infty$. 

The Hilbert space of functions on the sphere is then replaced by the Hilbert space of $(2j+1)\times (2j+1)$ matrices. The action of the new $\bm{n}$ operators is represented by matrix multiplication by $J_i$ and the action of the angular momentum operators by the commutators $[ J_i, \bullet]$.


Here, we study the fuzzy sphere with $j = 1/2$, for which the local Hilbert space at each site is again four-dimensional --- the space of $2\times2$ complex matrices with inner product $\braket{\psi|\phi} \equiv \tr (\psi^\dagger \phi)$. In this case, we use $J_k = \sigma_k/\sqrt{3}$, i.e., the Pauli matrices with a normalization factor. We compute the Hamiltonian in the ortho-normal basis $\{|\mathcal{J}_i\rangle\} = \{ {i}/{\sqrt{2}} \;  \bbid, \sqrt{3/2}\:J_i\}$.
Following the construction in Ref.~\cite{Alexandru:2019ozf}, we write the kinetic term of the Hamiltonian on $|\mathcal{J}\rangle$ as a double commutator
\begin{equation}
    (H_{0}^{f})_{ab} = -\frac{1}{2}\langle \mathcal{J}_b |\nabla^2|\mathcal{J}_a\rangle = \frac{\kappa}{2}\sum_{i=1}^{3} \tr (\mathcal{J}_b^{\dagger}[J_i,[J_i, \mathcal{J}_a]]);
\end{equation}
with a normalization constant $\kappa$. In fact, if $\kappa = j(j+1) = 3/4$, then we find that $H_0^f = \text{diag}(0, 1, 1, 1)$ reproduces the first four Laplacian eigenvalues on the sphere exactly, as in the truncated angular momentum model. The matrix elements for the fuzzy sphere coordinates $J_k$ in this basis are
\begin{equation}
    (j_{k})_{ab} = \langle \mathcal{J}_a |J_k|\mathcal{J}_b\rangle = \tr (\mathcal{J}_a^{\dagger}J_k\mathcal{J}_b).
\end{equation}
These $4\times4$ matrices are explicitly 
\begin{equation}
    j_1 = \frac{\bbid \otimes \sigma_2}{\sqrt{3}},\:\:\:j_2 = \frac{\sigma_2\otimes\sigma_3}{\sqrt{3}},\:\:\:j_3 = \frac{\sigma_2\otimes\sigma_1}{\sqrt{3}}.
\end{equation}
The fuzzy truncated theory Hamiltonian is thus given by
\begin{equation}
     \label{Hamiltonian_fuzzy} H_{F} = \eta g^2\sum_{x=1}^N H^{f}_{0}(x) \pm \frac{3\eta}{4g^2}\sum_{x=1}^{N-1}\sum_{k=1}^3 j_{k}(x) j_{k}(x+1).
\end{equation}
This form resembles the Hamiltonian with the $\ell_\text{max}$-truncation in Eq.~\eqref{Hamiltonian}. Furthermore, this Hamiltonian is also symmetric under $O(3)$ rotations~\cite{Alexandru:2019ozf}, and we use open boundary conditions. However, for this model the anti-ferromagnetic and ferromagnetic phases are distinct~\cite{Alexandru:2021xkf}. Here, we only consider the \textit{anti}-ferromagnetic phase since the ferromagnetic case was shown not to reproduce the $\sigma$-model~\cite{Alexandru:2021xkf}.   

\section{Methods}
\label{sec:methods}

To check whether the truncated models are in the same universality class as the full theory, we compute the energies of the lowest-lying states in a finite box 
as a function of the volume. For large enough boxes (i.e., $L \gg 1/m$), the volume shift in the energy is determined by two-particle elastic scattering phase shifts at momenta of order $\sim 1/L$ ~\cite{Luscher:1985dn}. The same shift in small boxes ($L\ll 1/m$) probes the ultraviolet behavior of the theory.

The MPS formalism we use here to analyse the different qubitizations is much more efficient when using open boundary conditions, so their use will allows us 
to probe more deeply into the ultraviolet regime than was permitted by periodic boundary conditions in Ref.~\cite{Alexandru:2021xkf}. In
Sec.~\ref{sec:mps} we summarise the MPS formalism and discuss how to determine correlation lengths.
The tuning of $\eta$ required to recover a relativistic theory is discussed in Sec.~\ref{sec:continuum}. Lastly, in Sec.~\ref{sec:mcmc}, we describe how we use Monte Carlo methods to obtain scaling results for the $\sigma$-model as a basis of comparison for each truncation.

\subsection{Matrix Product States}
\label{sec:mps}

 We use a variational MPS algorithm to compute low-lying eigenstates and eigenvalues of each Hamiltonian~\cite{White:1993}. The MPS ansatz for states of a $N$-site system is
\begin{equation}
    \label{MPS} |\Psi\rangle = \sum_{a_1, \ldots, a_N} A_1^{a_1}\cdots A_N^{a_N}|a_1,\ldots,a_N\rangle
\end{equation}
where $a_n = 1,\ldots, p$ (spanning the dimension $p$ of the local Hilbert space), $|a_1,\ldots,a_N\rangle$ are basis states, and $A_n^{a_n}$ are $D_{n-1}\times D_n$ matrices. In the open boundary case, the ends of the MPS chain $A_1^{a_1}$ and $A_N^{a_N}$ are $1\times D_1$ and $D_{N-1} \times 1$, i.e., row and column vectors respectively. 

This ansatz can describe any state exactly, if we allow these ``bond'' dimensions $D_n$ to grow to $p^{\mathrm{min}(n,N-n)}$
~\cite{Schollw_ck_2011}. 
Expressing arbitrary $N$-site states in a MPS form relies on iterative singular value decompositions (SVDs), and the ranks of the $A_n^{a_n}$ matrices may grow by $p$ at every step towards the center of the chain, in general. 
However, for states with more limited entanglement, as is the case for the ground states of gapped systems, we can produce a very good MPS approximation with relatively small matrix dimensions. 
In practical calculations, the dimensions of the MPS matrices are determined dynamically, based upon the singular values that appear in the SVD, and they are capped to a chosen maximum $D$, so that $D_n \leq D$. 
Also, while the MPS approximation generically becomes exact in the limit that the MPS space approaches the size of the full Hilbert space of the $N$-site system $p^N$ (i.e., $D$ approaches $p^{\lfloor N/2\rfloor}$), low-energy observables follow an area law and also converge quickly in $D$ for gapped systems. 
In particular, the ground state $|\Psi_0\rangle$ and its energy $a \hat E_0$ can be obtained by iteratively minimizing the expectation value of $H$ with respect to the $A_n^{a_n}$~\cite{Or_s_2014}. Excited states $|\Psi_k\rangle$ and energies $a \hat E_k$ are obtained via a similar algorithm, but with the additional constraints that the excited states are orthogonal to the lower-lying states $\langle \Psi_j|\Psi_k\rangle_{j<k} = 0$. 

\begin{figure}[t]
\includegraphics[width=\columnwidth]{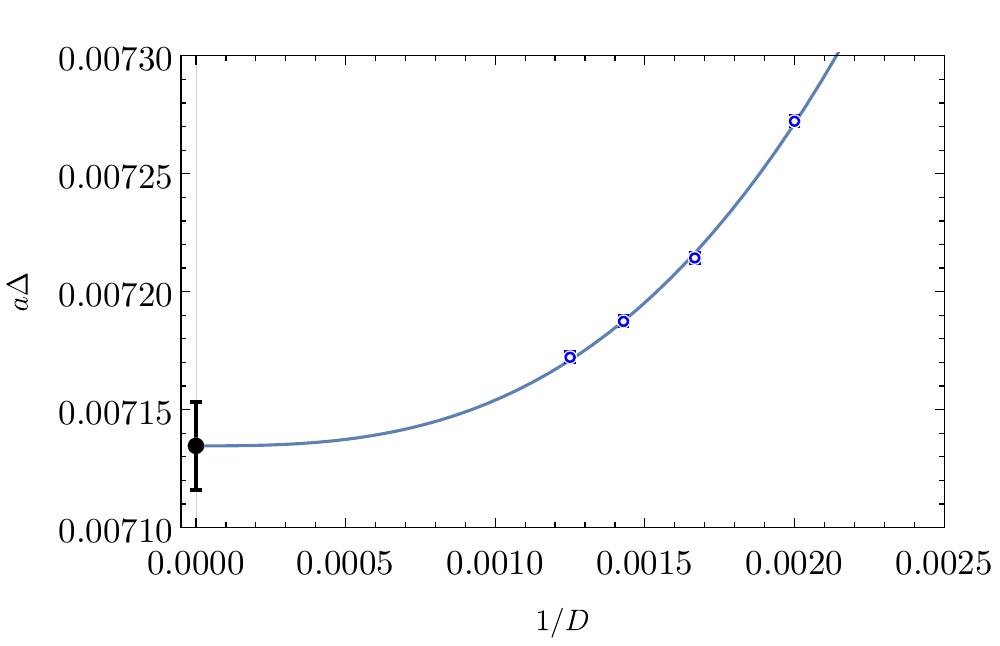}
\caption{Energy gap for the anti-ferromagnetic fuzzy model for $g^2=0.53$ and $N=800$ as a function of the MPS bond dimension $D$ together with the $D\to\infty$ extrapolation. The error bar on the extrapolation is determined by the distance between the value obtained from our largest choice of $D$ and the extrapolated value.}
\label{fig:Dextr}
\end{figure}

The systematic error due to the bond dimension cutoff $D$ constitutes the main source of error of the MPS algorithm. As such, final estimates for measured quantities must be extrapolated. For the energy gap $a\Delta := a \hat E_1 - a \hat E_0$, we find that a power law $\Delta(D) = \Delta + A/D^B$ yields suitable fits for the range of $D$ we use. A typical fit and extrapolation in $D$ is presented in Fig.~\ref{fig:Dextr}. Following Ref.~\cite{Bruckmann:2019}, we define our error on the observable as half of the distance between the extrapolated value for $D\to\infty$ and the value determined by our largest-$D$ calculation.

The main motivation for considering open boundary conditions is that the computational cost for contracting the tensors in the MPS ansatz in Eq.~\eqref{MPS} is significantly lower than that of the periodic boundary condition: $\mathcal{O}(NpD^3)$ vs.~$\mathcal{O}(NpD^5)$~\cite{Pippan:2010}. This speed-up of $\mathcal{O}(D^2)$ becomes significant when probing closer to the continuum limit of our Hamiltonian models, where large bond dimensions 
are required to accurately find states of large systems with large correlation lengths. Using open boundary conditions allows us to study the models deeper into the ultraviolet regime than previous studies using periodic boundary conditions~\cite{Alexandru:2021xkf}. Furthermore, there are well-established and performance-optimized libraries for executing open boundary MPS algorithms, such as ITensor~\cite{itensor}, which we utilize. 

\subsection{Taking the Continuum Limit}
\label{sec:continuum}

Equipped with a method of obtaining lowest-lying eigenstates of our Hamiltonian lattice theories, we proceed by demonstrating how these results may be used to compare each Hamiltonian model with the $\sigma$-model. 
Specifically, we determine correlation lengths for a given set of parameters in each model in Sec.~\ref{sec:corr}, and prescribe a method to tune the overall normalization factor $\eta$ in Sec.~\ref{sec:param_tuning}.

\subsubsection{Spatial Correlation Lengths}
\label{sec:corr}

To take the continuum limit, we need to tune the parameters of the model to a critical point, where the correlation length diverges. For the models discussed in Sec.~\ref{sec:models}, we expect that this point corresponds to $g^2\rightarrow 0$, similar to the original $\sigma$-model. Motivated by this observation, we scan this region in our calculations.


We determine the spatial correlation length by computing the point-to-point correlation function of the field operator in the ground state  $|\Psi_0\rangle$; the correlation function is defined by
\begin{equation}
    C(x, y) = \langle \Psi_0 |\mathcal O(x) \mathcal O(y)|\Psi_0\rangle \zeta^{x-y}
\end{equation}
where $\mathcal O=y_3$ for the angular-momentum truncation and $\mathcal O=j_3$ for the fuzzy sphere. We set $\zeta=1$ for the ferromagnetic case and $\zeta=-1$ for the anti-ferromagnetic case to remove the possible alternating sign in this correlator. 
The ground state for open boundary conditions is not translationally invariant, and the walls distort the correlation function. 
Importantly, the boundary effects diminish exponentially with distance from the walls. 
To minimize such wall effects, for a given distance $r$, we compute $C(r) = C(x,y=x+r)$ using the two points $x$ and $y$ equally distanced from the center of the box.\footnote{For even $N$ and even values of $r$ this goal is not possible, as one of the points must be closer to the center of the box by one. For this case, we take the average over two setups with either $x$ or $y$ closer to the center.}

Once $C(r)$ is obtained, we perform a series of two-parameter fits of the correlator to its expected form in 1+1 dimensions:
\begin{equation}
    \label{expectedC} C(r) = A {K}_0((a m)r),
\end{equation}
where $A$ is an amplitude, $a m = 1/\xi$ is the inverse spatial correlation length, 
and ${K}_0$ is the zeroth-order modified Bessel function of the second kind. By fitting the correlator on an interval $[x_0,x_0+w]$ for a window size $w$, one expects the extracted fit parameter $a m$ as a function of $x_0$ to form a plateau at large distances, where the correlator is a pure Bessel function. 

\begin{figure}[t]
\includegraphics[width=\columnwidth]{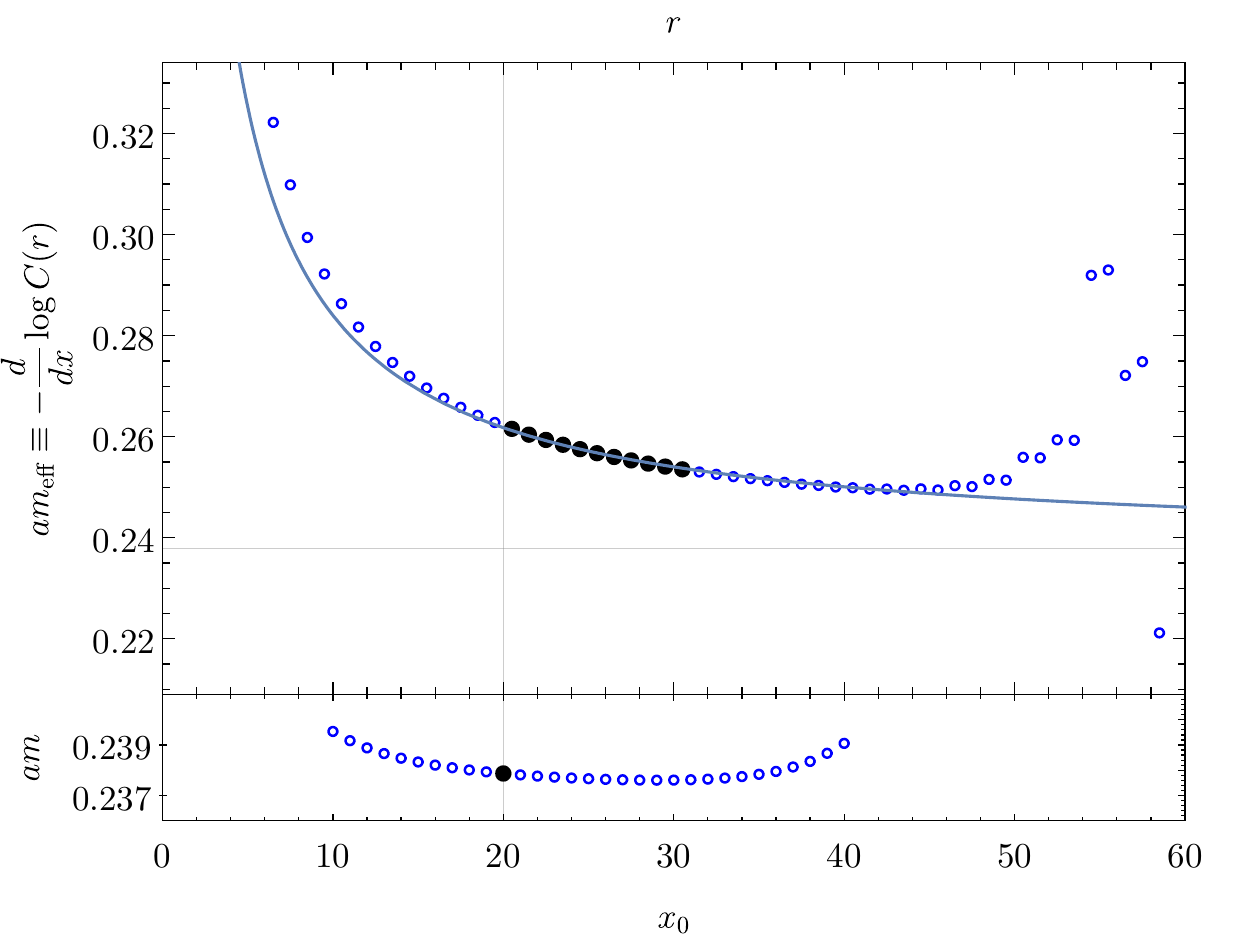}
\caption{A typical ground state correlator and the mass extracted from the fit window with the left edge at position $x_0$ along the chain. 
Specifically, we show results from 
the anti-ferromagnetic fuzzy model at $g^2=0.75$, $N=60$, and $D=800$. 
The top plot shows the {\it effective mass} (the logarithmic derivative, for the sake of visibility), and the bottom plot shows the mass extracted from the fit 
to the form in Eq.~\eqref{expectedC}. The black points represents the particular fit range used (above) and the value of $am$ extracted for that range (below). 
As $x_0$ approaches $40$, the fit range starts including data points close to the wall, and so the value of the mass extracted from the fit increases.
}
\label{fig:correlator}
\end{figure}

This procedure is illustrated in Fig.~\ref{fig:correlator}. 
In the top plot, we compare the correlator with the fitted function by plotting the effective mass, i.e., its logarithmic derivative. 
For a purely exponential correlator, the effective mass should have a plateau; 
however, the data clearly indicates that the correlator never approaches a pure exponential. On the other hand, the Bessel function form fits the data quite well, at least asymptotically. 
The bottom plot identifies the fitted value for $am$ as a function of the fit window. This mass estimate varies significantly less as we vary the fit window, compared to the effective mass above, and it develops a plateau for large $x_0$ values. 

The correlation function also suffers from finite-$D$ effects, as the correlator at large distances converges more slowly. 
Our fitting strategy is the following: we compute $m(D)$ as the minimum value extracted from fitting the correlator over all fit ranges.
The mass is extrapolated using a power law, $m(D) = m + A/D^B$, just as with the energy gaps in Sec.~\ref{sec:mps}. 
The error on the extrapolation is also defined in the same fashion: $\epsilon_m = [m - m(D_{\mathrm{max}})]/2$. 

\subsubsection{Parameter Tuning}
\label{sec:param_tuning}

The Hamiltonians for the truncated systems of Sec.~\ref{sec:models} are defined only up to a normalization factor $\eta$. This positive normalization factor does not affect the ground state of the system or its properties, such as the correlation length. Moreover, this factor only changes the energies (linearly). 
We note that this normalization is not specific to the truncated models; it is required for the original $\sigma$-model Hamiltonian as well~\cite{Shigemitsu:1981,Bruckmann:2019}. 
In a relativistic theory, the inverse correlation length $\xi=1/m $ is given by the  mass of the particle, which in the infinite volume is given by $\Delta$. We can enforce this relation by choosing
$\eta$ so that
\begin{equation}
\eta(g^2) = \frac{am(g^2)}{a \Delta(g^2)}
\end{equation}
where $\Delta(g^2)$ is the energy gap, $\Delta = \hat E_1 - \hat E_0$, at a given $g^2$, in the infinite-volume limit, computed from a Hamiltonian with $\eta$ set to 1. The renormalized energies are then $E_k := \eta \hat E_k$, and in particular, this definition guarantees that the renormalized energy gap in infinite volume $\eta a\Delta$ equals $am(g^2)$. In other words, the continuum limit is obtained by sending $g^2 \rightarrow 0$ along a line in $(g^2,\:\eta$)-space such that the infinite-volume energy gap coincides with the inverse spatial correlation length.

We can estimate  $\eta(g^2)$ nonperturbatively by obtaining estimates for the spatial correlation length $\xi$ and infinite volume gaps. In particular, we estimate $\xi$ in the manner described in Sec.~\ref{sec:corr}. Additionally, to obtain an infinite-volume gap $\Delta$, we obtain the energy gap at various lattice sizes $N$ and perform an extrapolation in $N$. 

For periodic boundary conditions \cite{Alexandru:2021xkf}, one could use Lüscher's formula for the one-particle finite-volume corrections \cite{Luscher:1986pf} to fit the gaps to known finite-volume behavior and extrapolate to the infinite volume. 
The finite-volume behavior depends on the scattering phase shifts of the theory, and they are exactly known for the $O(3)$ $\sigma$-model~\cite{1978NuPhB.133..525Z}.

On the other hand, for open boundary conditions, L\"uscher's original formulation does not apply. Instead, we expect that the finite-volume gap of an open boundary system can differ from the infinite-volume gap by a power of  the inverse volume. 
To motivate this claim, we observe that, perturbatively, the finite-volume change in a particle mass is given by the difference between loop diagrams computed in infinite and finite volumes or, in momentum space, by integrals or discrete loop sums. For the one-particle irreducible diagrams relevant to the mass, this difference behaves asymptotically like an exponential $\sim e^{-m L}$ when periodic boundary conditions are used. Open boundary conditions, on the other hand, eliminate the zero-mode $p=0$ that contributes a power of $1/L$.

\begin{figure}[t]
\includegraphics[width=\columnwidth]{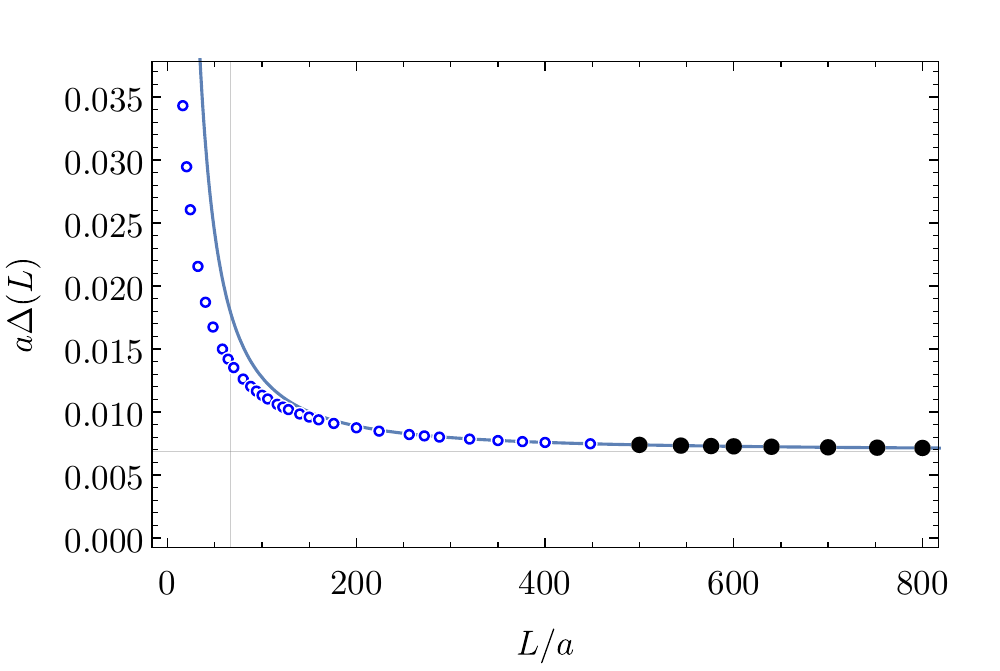}
\caption{A typical infinite volume extrapolation for the single-particle energy gap, as described in Sec.~\ref{sec:continuum}. 
Specifically, we show results from 
the anti-ferromagnetic fuzzy model with $g^2=0.53$. The solid (black) points in particular represent the data used in the fit range, and the curve in the result of this fit. The horizontal line is the infinite-volume result from the fit, while the vertical line corresponds to $mL=1$. The error bars for the gaps are present in this plot, but they are smaller than the size of the symbols.}
\label{fig:infvolextr}
\end{figure}

Therefore, we extrapolate the energy gaps to infinite volume using a power law,
\begin{equation}
    \label{powerlaw} a\Delta(N) = a\Delta + \frac{A}{N^{B}}
\end{equation}
where $A$, $B$, and $a\Delta$ are fit parameters. Small exponential corrections to Eq.~\eqref{powerlaw} are ignored, since we consider only system sizes where these corrections are negligible. To ensure this restriction, we performed the fits only in the range $mL=N/\xi \gtrsim 5$. An example of such an infinite volume extrapolation is presented in Fig.~\ref{fig:infvolextr}. For the data included in this figure, the fit form describes the finite-volume energy gap data quite well for a large range of sizes, down to $N/\xi \sim 2$.

Additionally, in the anti-ferromagnetic fuzzy sphere model, there is evidence that Lüscher's formula~\cite{Klassen:1990ub} for the $O(3)$ $\sigma$-model captures the finite-volume behavior of periodic boundary systems well~\cite{Alexandru:2021xkf}. 
Consequently, we check for several cases that the infinite-volume gaps $a\Delta$ obtained by extrapolating open boundary, finite-volume gaps using Eq.~\eqref{powerlaw} are consistent (within errors) with those obtained by extrapolating periodic boundary, finite-volume gaps using Lüscher's formula (see Fig.~\ref{fig:corrlengths}).

\subsection{Monte Carlo}
\label{sec:mcmc}
The results obtained from the truncated models using MPS are compared to lattice Monte Carlo calculations of the original $\sigma$-model, using the same open boundary conditions. 
Numerical methods are used, since
the exact results obtained 
using the thermodynamic Bethe ansatz~\cite{Hasenfratz:1990zz,Balog:2003yr} 
are not available for the open boundary conditions, which are more convenient for the MPS formalism.
 
We use the lattice action given by
\beq
S = - \beta \sum_{t,x} \Big[ \bo n(t,x) \cdot \bo n(t+1,x) + \bo n(t,x) \cdot \bo n(t,x+1) \Big]
\eeq
where $\bo n(t,x)$ are unit vectors. The boundary conditions are periodic in the time direction, but open in the space direction; on an $N_t \times N_x$ lattice, ``open'' in the $x$-direction means that no neighbor term couples $x=N_x$ and $x=0$. Further, we carried out simulations at $\beta=1.4,1.5,1.65,1.7,1.8$, using the Wolff cluster algorithm~\cite{Wolff:1988uh}. Ensemble sizes ranged from 10 million to 1.8 billion statistically independent $\bo n$-field configurations, as these large statistics were required to achieve precision comparable to the MPS method.

Estimates for the finite-volume energy gaps $E(L) := E_1(L) - E_0(L)$ in lattice units were obtained by computing time-slice correlators
\beq
C(t) = \frac{1}{N_x^2} \sum_{x,y} \langle \bo n(t,x) \cdot \bo n(0,y) \rangle
\eeq
and fitting them to an exponential $\exp(-E(L)t)$. Since the finite-temperature $T=1/(aN_t)$ effects are of the order $\exp(-m/T)$, we may keep these effects negligible by selecting $N_t$ such that $am N_t = m/T \approx 8$. 

We note that the boundary conditions make the extraction of $E(L)$ particularly challenging. As a consequence of the lack of translation invariance, it is impossible to project to zero-momentum states, while the spectrum of states present in the correlation function is quite dense. To make this point clear, note that the time-slice correlator for a free scalar field in a Dirichlet box is given by
\beq
C(t) = \sum_{n=\mathrm{odd}} A_n \mathrm{e}^{-\omega_n(L) t}
\eeq
where $\omega_n(L) = \sqrt{m^2 + p_n^2(L)}$, $p_n(L)=n \pi/L$, and $A_n = (2 L / n \pi)^2/\omega_n$. The lowest-order exponential in this correlator becomes dominant when $t \delta E \gg 1$, with $\delta E = \omega_3(L)-\omega_1(L) \approx 8\pi^2/(mL^2)$. To address this issue, we carried out simulations using $N_t \approx 8/(a\delta E)$.

\begin{figure}[t]
\includegraphics[width=\columnwidth]{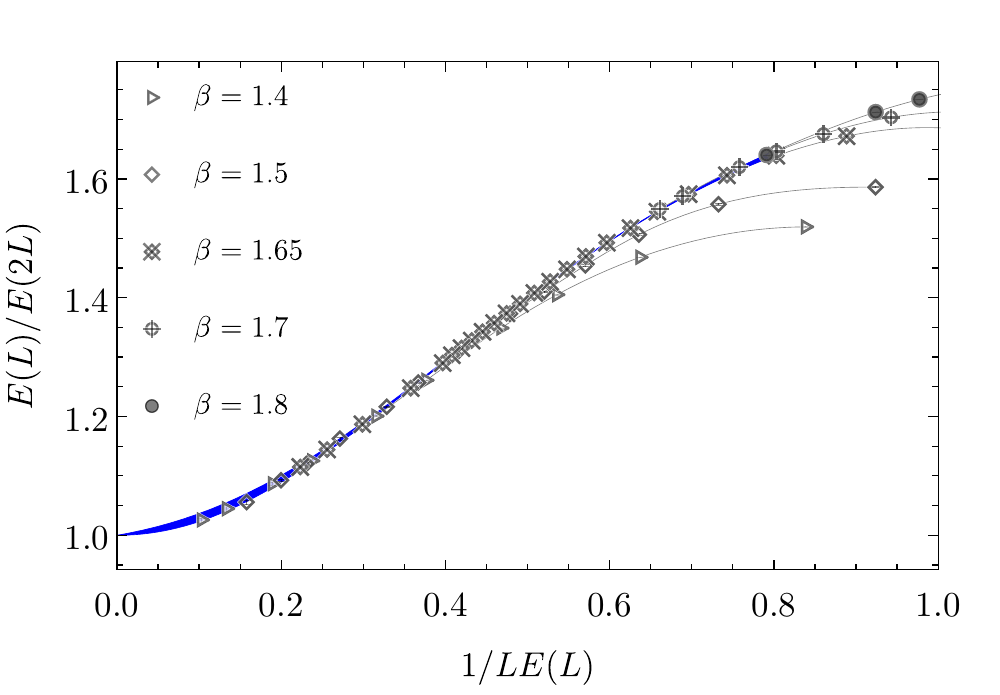}
\caption{Step-scaling curve for the $O(3)$ $\sigma$-model using open boundary conditions. The data points at each value of $\beta$ are joined using a second-order interpolation, to guide the eye. The solid (blue) band is the result of a fit to the envelope and represents our estimate for the step scaling curve in the continuum limit.}
\label{fig:mcscaling}
\end{figure}

From our simulations, we compute the {\it step-scaling curve}, which determines how the finite-volume energy responds to a doubling in the size of the system. The results for our Monte-Carlo simulations are presented in Fig.~\ref{fig:mcscaling}. As we increase $\beta$, the correlation length increases and the scaling curves approach a common envelope, which is the step-scaling curve in the continuum limit. It is clear from the plot that, for $1/L E(L)\leq 0.8$, at the level of the stochastic error-bars, the scaling curve in the continuum limit is already well described by the data produced at $\beta=1.65$. We fit these data to a simple parametrization, the ratio of two quadratic polynomials with free coefficient $1$. This fit is indicated in the figure by the blue band. This will be the reference data used to compare the truncated models against.

\section{Results}
\label{sec:results}

As a preliminary check for whether the truncated models are valid qubitizations of the $O(3)$ $\sigma$-model, we first compute inverse spatial correlation lengths $am$ at various values of $g^2$. A requirement for the model to be a valid qubitization is to have a critical point or continuum limit where the correlation length diverges $am \rightarrow 0$. To search for the critical points of our truncated models, we fit $am$ as a function of $g^2$ by its expected functional form~\cite{Hasenfratz:1990zz, Shigemitsu:1981}:
\begin{equation}
    \label{twoloop} am(g^2) = \frac{A}{g^2} \; \mathrm e^{-\frac{B}{g^2}} .
\end{equation}

\begin{figure}[t!]
\includegraphics[width=\columnwidth]{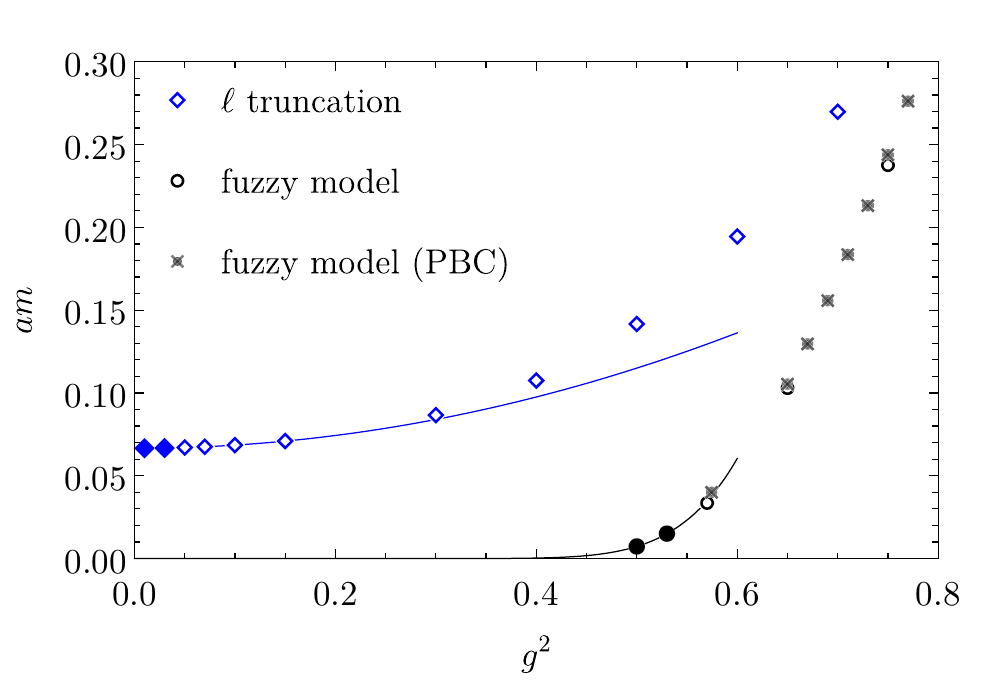}

\caption{The inverse correlation lengths $am$ plotted as a function of $g^2$ for each truncation model. The diamond points and overlapping curve (blue) are results from the $\ell_\mathrm{max}$-truncation, for which the correlation length stays finite ($\approx 15$ in lattice units) as $g^2$ approaches zero. The circular data points and overlapping curve (black) are results from the anti-ferromagnetic fuzzy model with a fit to Eq.~\eqref{twoloop}. The data suggest that the fuzzy model has a continuum limit at $g^2 \rightarrow 0$. The error bars are included but are too small to be seen at this scale. To obtain each curve, we fit simple models to the two lowest $g^2$ points (filled symbols). For the $\ell_\mathrm{max}$-truncation line we use a quadratic function in $g^2$. The cross data points (also black) are results from a previous study that employed periodic boundary conditions~\cite{Alexandru:2021xkf}.}
\label{fig:corrlengths}
\end{figure}

Figure~\ref{fig:corrlengths} presents these calculations for both models. We find that the anti-ferromagnetic fuzzy model gives $am(g^2)$ results which fit Eq.~\eqref{twoloop} well, suggesting a continuum limit indeed exists as $g^2 \rightarrow 0$. In contrast, the correlation length of the $\ell_\mathrm{max}$-truncated model remains finite in this region, yielding strong evidence that this model does not have a continuum limit and therefore is not in the same universality class as the $\sigma$-model. We also include the results for $am(g^2)$ as computed for the fuzzy model with periodic boundary conditions~\cite{Alexandru:2021xkf}; as expected, the results agree, which confirms the methodology used to extract these masses.

\begin{figure*}[t!]
\includegraphics[width=0.49\textwidth]{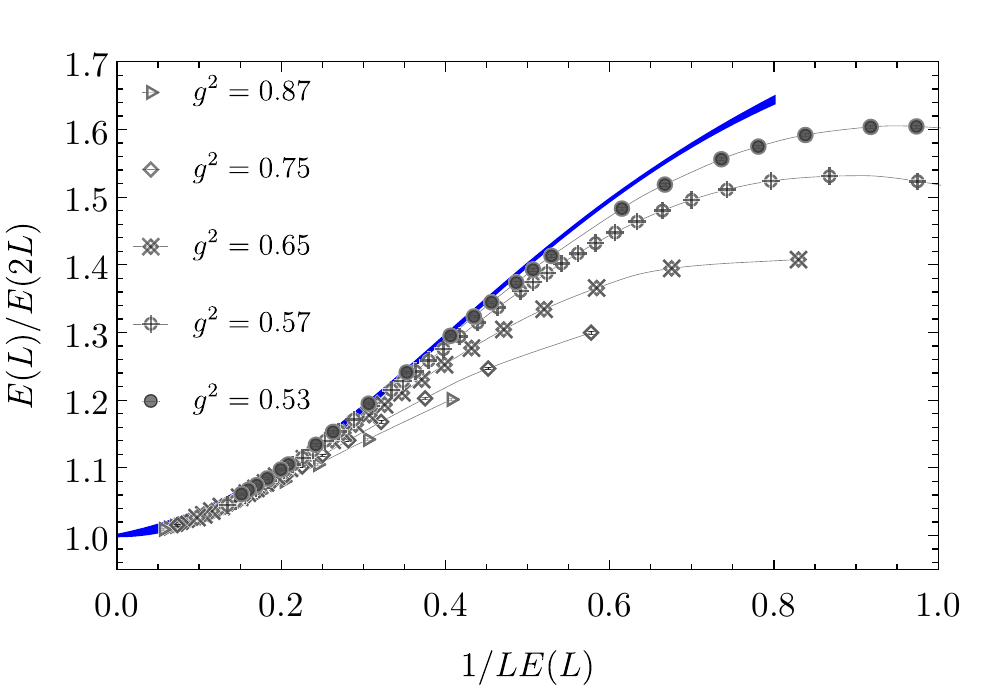}
\includegraphics[width=0.49\textwidth]{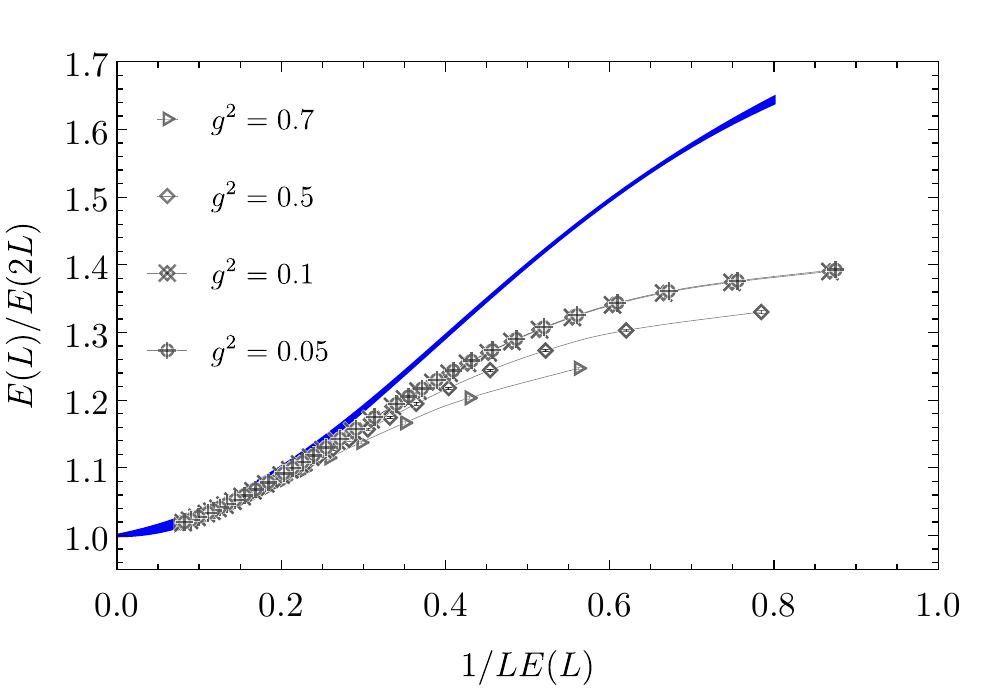}
\caption{\small{Step-scaling curves for the anti-ferromagnetic fuzzy model (left) and $\ell_\mathrm{max}$-truncation (right), compared with the $O(3)$ $\sigma$-model scaling curve computed from Monte Carlo simulations with \textit{open} boundary conditions (the blue band). Curves are interpolated between points with constant $g^2$. Rightmost points on each curve correspond to $(N,2N)$ pairs for the smallest lattice sizes ($N \sim 4$), while leftmost points correspond to the largest lattices (the sizes of which varied for different $g^2$). Points computed with the same $g^2$ are interpolated to make the envelope of the curves clear. \label{fig:caracciolo}}}
\end{figure*}

A further test of these models, over a wide range of energy scales, is to compare the step-scaling curves~\cite{Caracciolo:1994, Caracciolo:1995} to the full $\sigma$-model. Here, each curve is a plot of $E(L)/E(2L)$ as a function of $1/(LE(L))$. In the continuum limit, the curve should be the same for all models of the same universality class. 
In~\fig{fig:caracciolo}, we plot the zero-temperature scaling curve of the $O(3)$ $\sigma$-model determined in the previous section against the scaling curves for the angular momentum truncations with $\ell_{\text{max}} = 1$ and anti-ferromagnetic fuzzy models. 
For clarity, rightmost points on each curve correspond to the smallest lattice sizes ($N \sim 4$), while leftmost points correspond to the largest lattices (the sizes of which varied for different $g^2$). 

For both models, all constant-$g^2$ curves converge toward the Monte Carlo curve in the infrared regime (leftmost points), demonstrating a limited universality holding for the long-distance properties of each theory. 
On the other hand, we expect that truncated models in general will exhibit non-universal behavior as the scale of their lattice spacing is approached (namely, small $N$ for any fixed $g^2$). Indeed, rightmost points on each $g^2$-constant curve typically are far from the full-model curve. 
Note that we have seen similar behavior in Monte Carlo simulations of the lattice $\sigma$-model. However, we observe that, as $g^2$ decreases, the curves for the anti-ferromagnetic fuzzy model overlap with the full model curve further into the ultraviolet region, suggesting that more of the high-energy physics of the $\sigma$-model is reproduced. For the $\ell_\mathrm{max}$-truncation, as $g^2$ decreases there is a trend towards the full model curve, but this trend stops as the correlation length stops changing, consistent with the data in Fig.~\ref{fig:corrlengths}.

We note here that, since we used open boundary conditions, we were able to explore much larger correlation lengths and show that the trend toward the full model curve continues up to correlation lengths $1/am \approx 66.5$, providing stronger evidence that the anti-ferromagnetic fuzzy model and $\sigma$-model are likely in the same universality class.

\section{Discussion}
\label{sec:conclude}

We studied two different truncations of the field space of the 1+1-dimensional $O(3)$ $\sigma$-model, both preserving the symmetries of the full model. The first is an expansion of functions on $\mathcal{S}^2$ in spherical harmonics with an angular momentum cutoff $\ell_{\text{max}}$, such that the full Hilbert space of complex functions on a sphere is recovered as we take $\ell_{\text{max}} \rightarrow \infty$. Here, we explored the model with $\ell_\text{max}=1$. Another approach is to approximate $\mathcal{S}^2$ by a fuzzy sphere where the components of the position operator do not commute. Both models have the same dimension of Hilbert space at each site and would require the same number of quantum registers to implement the real time evolution. 

To assess whether these models lie in the same universality class as the $\sigma$-model, we calculated the finite-volume, single-particle energies in boxes with open boundary conditions. We used these results to compute the step-scaling curves and compare them with the results from the $\sigma$-model. 
The step-scaling curve for the fuzzy sphere truncation agrees with that of the $O(3)$ $\sigma$-model over a wide range of scales in the anti-ferromagnetic case, while the commutative sphere truncation fails to capture the $\sigma$-model behavior at energies beyond the deep infrared. 

It is possible that higher truncations $\ell_\mathrm{max} > 1$ will exhibit more desirable scaling properties than the one considered here did. For example, to converge to the expected gap for $\beta=1.8$ in the lattice $\sigma$-model, where the correlation length is similar to the largest one considered for the fuzzy model in this study, we need $\ell_\text{max} \gtrsim 3$~\cite{Bruckmann:2019}. 
However, we point out that for quantum simulations---the main motivation for developing these truncations---these truncations quickly become impractical. To implement a single time-step evolution, $\exp(-i H\delta t)$, the number of quantum gates increases very quickly with $\ell_\text{max}$. Using methods presented in Ref.~\cite{Murairi:2022zdg}, we can design quantum circuits to evolve a two-site system: for $\ell_\text{max}=1$, $2$, and $3$ we need $60$, $3826$, and $11826$ CNOT gates, respectively~\cite{Murairi:comm}. Hence, this strategy of removing the truncation by repeating calculations with increasing $\ell_\text{max}$ is not feasible. We stress that this rapid increase in complexity of the quantum circuits with the size of the Hilbert space is a generic feature, not particular to the $\sigma$-model. As such, for quantum simulations to be feasible for bosonic field theories, small size qubitizations are crucial.

For the fuzzy sphere truncation in the anti-feromagnetic case, we find that the finite-volume energy for the single-particle states agrees very well with the expectations from the $\sigma$-model. This observation provides further evidence that this model is in the same universality class. Moreover, this model requires only two qubits per site and has a very compact time evolution circuit~\cite{Alexandru:2019ozf}. We note that another qubitization was proposed using spin-ladder operators that has the same number of qubits per site~\cite{Singh:2019uwd}. For the $\sigma$-model, these truncations are likely to be the most economical ones for quantum simulations.

For other quantum field theories that include bosonic degrees of freedom --- quantum chromodynamics being a prime example --- designing appropriate qubitizations is crucial for simulating them on quantum computers. One important ingredient is to preserve the symmetries of the original model as much as possible. However, it is clear that this requirement is not sufficient, and discovering what other design principles are required is an important research task.

\begin{acknowledgments}
This work was supported in part by the U.S. Department of Energy, Office of Nuclear Physics under Award Number(s) DE-SC0021143,  and DE‐FG02‐93ER40762, and DE-FG02-95ER40907. The numerical results were produced in part with resources provided by the High Performance Computing Cluster at The George Washington University, Research Technology Services.
\end{acknowledgments}

\bibliographystyle{apsrev4-1}
\bibliography{fuzzy-mps.bib}

\end{document}